\documentclass[prd,twocolumn,nofootinbib]{revtex4}
\usepackage{amsfonts}
\usepackage{amsmath}
\usepackage{epsfig}

\def\beq{\begin{equation}}
\def\eeq{\end{equation}}
\def\bea{\begin{eqnarray}}
\def\eea{\end{eqnarray}}
\def\bq{\begin{quote}}
\def\eq{\end{quote}}
\def\simlt{\stackrel{<}{{}_\sim}}
\def\simgt{\stackrel{>}{{}_\sim}}

\newcommand{\YY}[1]{{Y}_\nu^{#1}}

\begin{document}

\title{Quasi-degenerate neutrinos and leptogenesis from $L_\mu-L_\tau$}
\date{\today}
\author{E.\ J.\ Chun}
\affiliation{Michigan Center for Theoretical Physics, Department of Physics, University of Michigan, Ann Arbor MI 48109, USA}
\affiliation{Korea Institute for Advanced Study, 207-43 Cheongryangri-dong, Dongdaemun-gu, Seoul 130-012, Korea}
\author{K.\ Turzy\'nski}
\affiliation{Department of Physics, University of Michigan, Ann Arbor MI 48109, USA}

\begin{abstract}
 We provide a framework for quasi-degenerate
 neutrinos  consistent with a successful leptogenesis,
 based on the $L_\mu-L_\tau$ flavor symmetry and its breaking pattern.
 In this scheme, a fine-tuning is needed to arrange the small solar neutrino
 mass splitting.  Once it is ensured,
 the atmospheric neutrino mass splitting and the deviation from
 the maximal atmospheric mixing angle $\Delta
 \theta_{23}$ are driven by the same symmetry breaking parameter $\lambda\sim 0.1$,
 and the reactor angle $\theta_{13}$ is predicted to be slightly smaller than $\lambda$
 while the Dirac CP phase is generically of order one.
 Given that the pseudo-Dirac nature of right-handed neutrinos is protected
 from the flavor
 symmetry breaking,
 a small mass splitting can be generated radiatively.
 For moderate values of $\tan\beta\sim 10$,
 this allows for low-scale supersymmetric leptogenesis,
 overcoming a strong wash-out effect
 of the quasi-degenerate light neutrinos
 and evading the gravitino overproduction.

\end{abstract}

\preprint{MCTP-07-07}

\maketitle

\section{Introduction}

Thanks to an impressive progress made in neutrino oscillation
experiments, we have fairly good information of the low-energy
observables like neutrino mass differences and mixing angles
\cite{Strumia06}.
The least known parameter is the so-called reactor angle
$\theta_{13}$. A measurement of this angle and a study of
CP violation in
neutrino oscillation is one of major tasks in the next neutrino
oscillation
experiments. These endeavors cannot, however, reveal
what the absolute neutrino
mass scale is.
A future determination of this feature would be a key element in
exploring the origin of neutrino mass, which clearly lies
beyond the Standard Model (SM).
Among all still allowed possibilities, the scenario
of quasi-degenerate neutrinos is interesting,
as it can be confirmed or disproved in
the future neutrino-less double beta decay experiments or in the
cosmological observations of the cosmic microwave radiation and
the large scale structure of the universe \cite{Strumia06}.

One of the most fascinating connections between neutrino physics
and cosmology would be a possible explanation of the baryon
asymmetry of the Universe,
$\eta_B\equiv(n_B-n_{\bar{B}})/n_\gamma=6.15(25)\times 10^{-10}$
\cite{wmap3} through leptogenesis \cite{Fukugita86} (see also
\cite{Buchmuller04} for a review of subsequent developments),
which is linked with the neutrino masses and mixing originating
from the seesaw mechanism \cite{seesaw}. Recent studies of
leptogenesis revealed a meaningful constraint on the scale of the
heavy right-handed neutrino mass $M$ at which the baryon asymmetry
is generated. Under the assumptions of a hierarchy in the masses
of the heavy right-handed neutrinos and a CP phase of order one in
their decay, $M\simgt10^8-10^9$ GeV is required to account for the
baryon asymmetry of the Universe, if the inverse-decay of this
right-handed neutrino is negligible \cite{Davidson02}. In case of
the quasi-degeneracy for low-energy neutrinos, the resulting
leptonic CP  asymmetry is suppressed by a strong inverse-decay
effect coming from larger neutrino Yukawa couplings, and as a
result, one needs to increase the scale $M$ by a factor of
$\sim10^{2-3}$ compared to the above value. Such a high
leptogenesis scale $M$ sets a lower bound on the the reheating
temperature after inflation, which may endanger the successful
prediction of the primordial nucleosynthesis due to gravitino
overproduction \cite{gravitino}.

Of course, the above-mentioned constraint on $M$  is
model-dependent. For instance, nearly mass-degenerate
right-handed neutrinos can lead to an increase in the asymmetry
\cite{resonance}. In fact, the quasi-degeneracy of the low-energy
neutrinos could be a consequence of that of the high-energy
right-handed neutrinos. An extreme possibility along this line is
to have the right-handed neutrino mass difference comparable to
their decay rate $\Delta M \sim \Gamma$, which leads to the
leptonic CP asymmetry resonantly enhanced to its near maximum
value and thus the mass scale $M$ can go down to the TeV scale
\cite{Pilaftsis97}. An interesting way of realizing such a
resonant enhancement is to invoke a radiatively induced mass
splitting through the renormalization group running from the
flavor scale to the mass scale $M$ \cite{Turzynski04,Branco05}.
Such radiative resonant leptogenesis has also been studied in the
context of minimal flavor violation \cite{mfv} and $\mu$-$\tau$
symmetry \cite{mutau}. An almost exact degeneracy requires a
theoretical justification; in a
flavor model of neutrino masses and mixing, an (nearly) exact
degeneracy of the singlet right-handed neutrino sector can be a
consequence of the flavor symmetry and should also be protected
from its breaking effect \cite{Branco05}.

In this work, we have taken the viewpoint that, since
baryogenesis {\em via} leptogenesis is a theoretically elegant
explanation of the baryon asymmetry of the Universe, a requirement
of successful leptogenesis (with a low reheating temperature to
avoid the gravitino problem) can be added to the list of
phenomenological constraints that a neutrino seesaw mass model
should observe\footnote{One should, however, remember that, unlike
the results of purely empirical studies, this constraint
introduces a strong theoretical prior, as it heavily relies on
several presently unverifiable assumptions: (i) that the neutrino
masses are generated in the seesaw mechanism, (ii) that the baryon
asymmetry of the Universe is generated through leptogenesis, and
(iii) that the mechanism of supersymmetry breaking predicts
the gravitino mass in the range potentially dangerous for
primordial nucleosynthesis. There exist models abandoning some of
these assumptions.}. We illustrate this point by investigating the
properties of a quasi-degenerate neutrino mass model based on the
$L_\mu-L_\tau$ flavor symmetry, which also provides a  successful realization of
the radiative resonant leptogenesis. The $L_\mu-L_\tau$ flavor
symmetry is motivated by the fact that the symmetry-preserving
right-handed neutrino mass term, $M N_\mu N_\tau$, naturally leads
to a maximal mixing required for the atmospheric neutrino
oscillation, $\theta_{23} = \pi/4$ \cite{lmlt0,lmlt1,lmlt1a,lmlt2}. Note also
that such a pseudo-Dirac structure of in the $\mu$-$\tau$ sector
implies an exact degeneracy for two right-handed neutrinos $M_2 =
M_3 =M$. As a consequence of it, the resulting low-energy neutrino
mass pattern is required to be quasi-degenerate, and a fine-tuning
has to be introduced to arrange a small mass splitting for the
solar neutrino oscillation as we will discuss in detail. We will
analyze how the atmospheric and solar mass splitting can arise in
connection with a small reactor angle and a large solar angle from
the flavor symmetry breaking which introduces small complex order
parameters $\lambda_i$, suppressed by a factor $\lambda ={\cal O}(0.1)$
with respect to the symmetry preserving ones. Such a flavor
symmetry breaking effect can be exempt in the right-handed
neutrino mass matrix by assigning an additional discrete symmetry,
as a result of which the resonant leptogenesis can naturally
explain the observed baryon asymmetry of the universe for
$\tan\beta \sim 10$ (thereby partially compensating the
above-mentioned fine-tuning). Our scheme predicts generically
order-one CP phases for the neutrino oscillation and leptogenesis
which are unrelated to each other.

\section{Quasi-degenerate neutrino mass model}

\subsection{General remarks}

Let us write down the  Lagrangian with three right-handed
neutrinos $N$ as
 \beq
 {\cal L} = N \YY{} L H_2 + \frac{1}{2} N {M} N + h.c.
 \eeq
which leads to  the seesaw mechanism explaining the smallness of
the neutrino masses:
 \beq
 {m}_\nu = -\langle H_2 \rangle^2
\YY{T} {M}^{-1} \YY{} \label{seesaw}
 \eeq
where ${m}_\nu$ is the mass matrix of the light neutrinos,
${M}$ is the Majorana mass matrix of the heavy right-handed
neutrinos and $\YY{}$ is the matrix of the neutrino Yukawa
coupling. The matrix ${m}_\nu$ can be diagonalized by a
unitary flavor transformation:
 \beq
 {V}_\nu^T
{m}_\nu {V}_\nu = \mathrm{diag}(m_1,m_2,m_3)
\label{eqdiag}
 \eeq
where we take $m_1,m_2,m_3$ to be {\it a priori} complex and the
neutrino mixing matrix  has a CKM-like form: 
\begin{widetext}
\beq {V}_\nu =
\left(
\begin{array}{ccc}
c_{12}c_{13} & s_{12}c_{13} & \tilde s_{13}^\ast \\
-c_{23}s_{12}-s_{23}c_{12}\tilde s_{13} & c_{23}c_{12}-s_{23}s_{12}\tilde s_{13} & c_{13}s_{23} \\
s_{23}s_{12}-c_{23}c_{12}\tilde s_{13} & -s_{23}c_{12}-c_{23}s_{12}\tilde s_{13} & c_{13}c_{23}
\end{array}
\right) \, .\label{ckmlike}
 \eeq
\end{widetext}
Here $\tilde s_{13}=s_{13}e^{\imath\delta}$, 
and $s_{ij}$, $c_{ij}$ stand for $\sin\theta_{ij}$,
$\cos\theta_{ij}$, respectively. The experimental constraints on
the dimensionless neutrino observables can be summarized as (see,
e.g., \cite{Strumia06}):
\begin{eqnarray}
&
\frac{|m_2|^2-|m_1|^2}{|m_3|^2-|m_1|^2} = 0.032(3)\, ,  &\nonumber\\
&
\sin\theta_{13} = 0.00(5)\, , &\nonumber\\
&
\tan^2\theta_{12}= 0.45(5)\, ,  &\nonumber\\
\label{expval}
&
\tan^2\theta_{23} = 1.0(2) \, . 
\end{eqnarray} 
The challenge of
building a neutrino flavor model consists in reproducing this
peculiar observed pattern of the neutrino mass squared differences
and two large and one small mixing angles.

\begin{figure}
\begin{center}
\hspace{-3cm}
\includegraphics*[height=9cm]{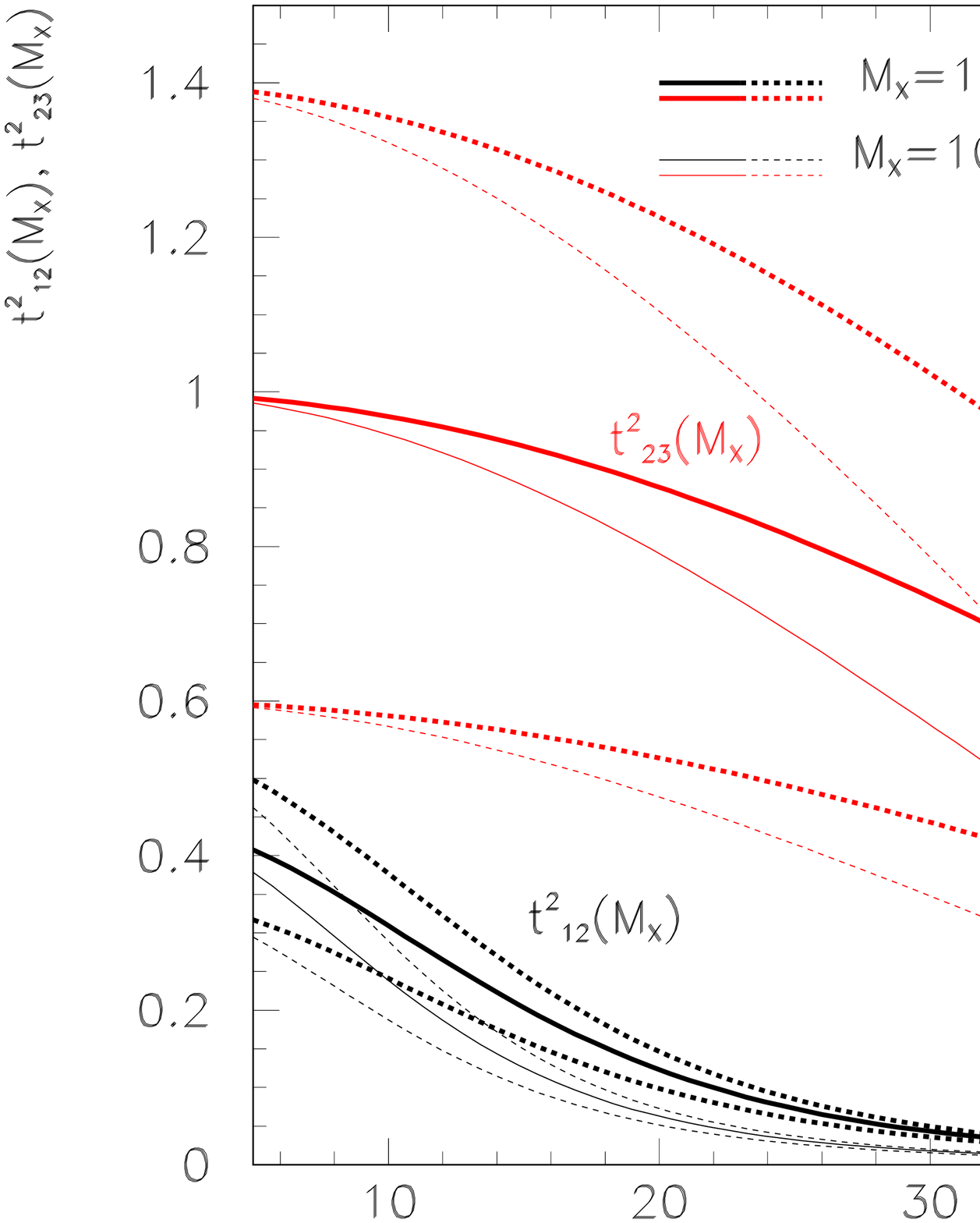}
\end{center}
\caption{Running $t_{12}(M_X)$ and $t_{23}^2(M_X)$ for $m_1=0.1\,\mathrm{eV}$ at two scales $M_X=10^8\mathrm{GeV}$ and $10^{14}\mathrm{GeV}$. In the latter case, the neutrino Yukawa couplings corresponds to Casas-Ibarra matrix ${R}={1}$ and they are included into the RG equations at appropriate thresholds $M_{N_A}=(1.0,1.1,1.2)\times 10^8\mathrm{GeV}$. The central solid (outer dashed) lines correspond to central values (2$\sigma$ deviations) of the relevant observables. We also chose $s_{13}=0.075$. \label{figrge}}
\end{figure}

Writing Eqs.\ (\ref{seesaw})-(\ref{ckmlike}), we tacitly assumed
that they are valid at the low energy scales at which the neutrino
experiments are performed. In order to match these expressions
with the neutrino Yukawa couplings and the masses of the
right-handed neutrinos defined at the high scale at which the
right-handed neutrinos are integrated out, one needs to compute
quantum corrections which typically contain large logarithms due
to a vast difference in the energy scales. These large logarithms
can be conveniently summed up with the use of the renormalization
group (RG) technique \cite{Chankowski00}. The RG corrections to
the neutrino masses and mixing angles in the Supersymmetric
Standard Model are particularly large for the degenerate mass
spectrum with definite CP parities $(\mp,\mp,\pm)$ and for large
$\tan\beta$. In particular, for the overall neutrino mass scale
$\sim 0.1\,\mathrm{eV}$ and $\tan\beta\simgt15$, the RG
corrections normally drive $\sin2\theta_{12}$ towards a small
fixed-point value inconsistent with experimental constraints,
unless the neutrino mass model is very finely tuned at high scales
\cite{Chankowski01}. Here, we adopt a perspective that the
observed pattern of the neutrino masses and mixing does not
accidentally emerge from the RG corrections, but that it reflects
features of the underlying flavor model, and  therefore our scheme
fits better for $\tan\beta\simlt15$. 
We illustrate this point in Figure \ref{figrge}, where we plot
the running $t_{23}^2(M_X)$ and $t^2_{12}(M_X)$ as functions of
$\tan\beta$ for two values $M_X=10^8,\,10^{14}$ GeV.
The values of $t_{23}^2(M_X)$ and $t^2_{12}(M_X)$ corresponding
to $t^2_{23}$ and $t^2_{12}$ fixed at their best-fit values 
(2$\sigma$ deviations) at the low scale, as given in (\ref{expval}),
are shown as central solid (outer dashed) lines.
We also chose
$m_1=0.1\,\mathrm{eV}$, $s_{13}=0.075$ 
and $M_{N_A}=(1.0,1.1,1.2)\times 10^8\mathrm{GeV}$.
For such low masses of the right-handed neutrinos, the effects
of the contributions from the neutrino Yukawa couplings are negligible,
nevertheless, for $M_X=10^{14}$ GeV we include them into the RG
equations, choosing the texture corresponding to
the Casas-Ibarra matrix ${R}={1}$ \cite{casas01}
and `switching on' the relevant neutrino Yukawa coupling at the 
appropriate thresholds.
It is convenient to note that
the procedure of deriving the RG equations for the neutrino masses
and mixing angles allows maintaining an arbitrarily chosen phase
convention for the neutrino masses and the neutrino mixing matrix
\cite{Turzynski05}, and we shall utilize it to adhere to the phase
choice corresponding to Eq.\ (\ref{ckmlike}) throughout the entire
RG evolution.

\subsection{The $L_\mu-L_\tau$ flavor model}
\label{flamo}

Flavor structure of the Yukawa couplings of quarks and leptons are
often explained with the use of a Froggatt-Nielsen mechanism with
one or more flavons $\Phi_A$, i.e.\ scalar fields acquiring vacuum
expectation values (vevs), spontaneously breaking a beyond-SM
flavor symmetry, but coupling to the matter fields in a
symmetry-preserving manner. One of numerous attempts to address
the empirically determined pattern of the neutrino observables is
to postulate an approximate $L_\mu-L_\tau$ global $U(1)$ symmetry
in the lepton sector, which has a virtue of predicting almost
maximal atmospheric mixing from a pseudo-Dirac structure of the
right-handed neutrino mass matrix. In addition, we assume that
there is a discrete ${Z}_n$ symmetry in the lepton-flavon
sector. A full field content and charge assignment is given in
Table \ref{tabfla}.
The couplings allowed by symmetries give rise to the almost
maximal atmospheric mixing, while those arising by the spontaneous
breaking of $U(1)$ and thus flavor-scale suppressed allow
reproducing the remaining features of the neutrino masses, given
that certain constraints are fulfilled. The presence of
${Z}_n$ symmetry prevents the flavon vevs from contributing
to the mass matrix of the right-handed neutrinos, which shall turn
out to be important for the possibility of radiative resonant
leptogenesis. We would, however, like to stress that our goal
consist in exploring a phenomenologically motivated neutrino
flavor pattern which may provide successful leptogenesis rather
than in pretending that our construction is the ultimate model of
leptonic flavor.

\begin{table}
\begin{center}
\begin{tabular}{|c|ccc|ccc|cc|}
\hline
field & $L_e$ & $L_\mu$ & $L_\tau$ & $N_e$ & $N_\mu$ & $N_\tau$ & $\Phi_{\pm1}$ & $\Phi_{\pm2}$\\
\hline
$U(1)$ charge & 0 & $+1$ & $-1$ & 0 & $-1$ & $+1$ & $\pm1$ & $\pm2$ \\ 
${Z}_n$ multiplicity & 1 & 1 & 1 & 0 & 0 & 0 & $n-1$ & $n-1$ \\ 
\hline
\end{tabular}
\end{center}
\caption{Field content and the charge assignment in the lepton-flavon
sector. \label{tabfla}}
 \end{table}

The resulting neutrino Yukawa matrix and the Majorana mass matrix
of the right-handed neutrinos  are:
 \beq
\YY{} = \frac{1}{\langle H_2\rangle}
\left(
\begin{array}{ccc}
a & \lambda_1 & \lambda_2 \\
\lambda_6 & b & \lambda_3 \\
\lambda_5 & \lambda_4 & d
\end{array}
\right)
 \label{assumpt0}
\eeq
and
\beq
{M} =
\left(
\begin{array}{ccc}
X & 0 & 0 \\
0 & 0 & Y \\
0 & Y & 0
\end{array}
\right) \, . \label{assumpt}
 \eeq
The parameters $X$ and $Y$ consistent with the $U(1)$ symmetry are
of the same order of magnitude, and the same is true for $a,b,d$.
On the other hand, the flavor symmetry breaking parameters $\lambda_i$ are smaller than the latter as they arise from the flavon vacuum expectation values: $\lambda_{1,5} \propto \langle\Phi_{-1}\rangle/M_X$, 
$\lambda_{2,6} \propto \langle\Phi_{+1}\rangle/M_X$ and $\lambda_{3,4} \propto \langle\Phi_{\pm2}\rangle/M_X$ where $M_X$ is the flavor symmetry breaking scale. We take the flavor suppresion factors 
$\lambda_i/a$ of the order  $\lambda\sim\mathcal{O}(10^{-1})$. 
It follows
from the pseudo-Dirac structure the 2-3 sector of ${M}$
that two right-handed neutrinos are exactly degenerate in masses,
while the mass of the third right-handed can be slightly
different.

So far, we have not chosen any specific phase convention for the
right-handed neutrinos. We can use the transformations $N_1\to
e^{\imath\varphi_1}N_1$ and $N_{2,3}\to
e^{\imath\varphi_2}N_{2,3}$ to ensure that $X$ and $Y$ are real
and positive. The remaining phase redefinition $N_{2,3}\to
e^{\pm\imath\varphi_3}N_{2,3}$ leaves ${M}$ invariant, but
it changes phases in the second and third row of the neutrino
Yukawa matrix. We can also make the phase redefinitions of
the charged lepton doublets, $L_i\to e^{\imath\phi_i}L_i$
($i=e,\mu,\tau$).
First,
we can redefine the overall leptonic phase $\phi_e+\phi_\mu+\phi_\tau$
and the phase $\varphi_3$ so that $d$ is
real and positive, and $b$ is real and negative (these transformations
do not depend on the phase convention imposed by Eq.\ (\ref{ckmlike})).
The remaining freedom of the phase choice must be then utilized
to ensure that the neutrino mass matrix
is diagonalized with a matrix of the form
(\ref{ckmlike}). As we shall see, this will introduce some
consistency constraints.
These unphysical phases correspond to
the freedom of $\phi_e$ (allowing to set an arbitrary
phase to $a$)
and to the freedom of shifting $\varphi_3$,
$-\phi_\mu$ and $\phi_\tau$ by the same value; it would be a
symmetry of the neutrino Yukawa matrix, if the $L_\mu-L_\tau$
breaking were absent.

Given the form of the neutrino mass matrix from 
Eq.~(\ref{assumpt0})-(\ref{assumpt}),
the low-energy observables like neutrino mass splitting
and mixing angles can be explicitly calculated perturbatively
treating the small symmetry breaking entries as expansion
parameters. Using Eqs.\ (\ref{eqdiag}) and (\ref{ckmlike}), we can
expand the neutrino mass matrix around $s_{13}=0$ and
$\theta_{23}=\pi/4$ for an arbitrary $\theta_{12}$ as:
\beq
{m}_\nu =
{m}_\nu^{(0)}+{m}_\nu^{(1)}+
\ldots
\label{mnexp}
\eeq
In Eq.\ (\ref{mnexp}),
${m}_\nu^{(0)}$ is the
neutrino mass matrix in the limit $\lambda_i,\delta_a^{(n)}\to 0$
and ${m}_\nu^{(1)}$ accounts for
The $\mathcal{O}(\lambda)$ correction
to the neutrino mass matrix:
\begin{eqnarray}
{m}_\nu^{(1)} &=& 
\left. {m}_\nu \right|_{\begin{array}{l} s_{13}=0\\\theta_{23}=\pi/4\\ m_i=m_i^{(1)}\end{array}}  
+
\Delta s_{13}^{(1)}\left. \frac{\partial{m}_\nu}{\partial s_{13}} \right|_{\begin{array}{l} s_{13}=0\\\theta_{23}=\pi/4\\ m_i=m_i^{(0)}\end{array}}
+ \nonumber\\
&&
+\Delta \theta_{23}^{(1)}\left. \frac{\partial{m}_\nu}{\partial \theta_{23}} \right|_{\begin{array}{l} s_{13}=0\\\theta_{23}=\pi/4\\ m_i=m_i^{(0)}\end{array}}
\label{exp1}
\end{eqnarray}
where $\Delta s_{13}^{(1)}$ and
$\Delta\theta_{23}^{(1)}$ are corrections 
to the leading pattern of the neutrino
mixing, and $m_i^{(1)}$ are corrections to the eigenvalues
of the neutrino mass matrix.
It is straightforward to derive higher order terms of this expansion.
Now we shall compare the neutrino mass matrix decomposed as described above
with the
mass matrix resulting from (\ref{assumpt}) {\em via} (\ref{seesaw}).

At the leading order,  $\mathcal{O}(\lambda^0)$, we obtain a
neutrino mass matrix which has a pseudo-Dirac structure in the 2-3
sector. This picture can be extended to the first generation,
predicting an exactly degenerate mass spectrum, given that: 
\beq
\frac{a^2}{X} = \left(-\frac{bd}{Y}+\sum_{n}\delta_a^{(n)}\right) e^{\imath\alpha} \, ,
\label{ft1} 
\eeq 
where $\delta_a^{(n)} \sim\mathcal{O}(\lambda^n)$ 
are real and $\alpha\sim\mathcal{O}(\lambda)$.
Then we find that
$(m_1^{(0)},m_2^{(0)},m_3^{(0)}) = (-1,-1,+1)\times |b|d/Y$ the
atmospheric mixing is maximal, $s_{13}$ vanishes and the solar
mixing remains undetermined. Clearly, Eq.\ (\ref{ft1}) indicates
that our model requires a fine-tuning to describe the neutrino
masses and mixing correctly. We shall address the issue of actual
fine-tuning compared to other neutrino mass models in the
following section.

Let us turn to calculating $\mathcal{O}(\lambda)$ corrections to this result.
{}From now on, we shall further simplify our model by setting
$\lambda_1=\lambda_5=0$, which is ensured by the absence of the flavon field
$\Phi_{-1}$.   Such an assumption does not change
qualitative features of the model, while simplifying the following
formulae.
Comparing the sum and the difference of the 12 and 13 entries of 
${m}_\nu^{(1)}$ with the appropriate combinations of the
$\mathcal{O}(\lambda)$ entries of the neutrino
mass matrix resulting from (\ref{assumpt0}) and (\ref{assumpt}) 
{\em via} (\ref{seesaw}):
\beq
m_\nu =
\left(
\begin{array}{ccc}
\frac{a^2}{X} & \frac{\lambda_4\lambda_6}{Y} & \frac{\lambda_2a}{X}+\frac{\lambda_6d}{Y} \\
\ast & \frac{2\lambda_4b}{Y} & \frac{bd+\lambda_3\lambda_4}{Y} \\
\ast & \ast & \frac{2\lambda_3d}{Y}+\frac{\lambda_2^2}{X}
\end{array} 
\right) \, ,
\label{mnuour}
\eeq
(entries denoted by $\ast$ are given by symmetry of $m_\nu$)
we obtain:
\begin{eqnarray}
-\frac{a\lambda_2}{X}-\frac{d\lambda_6}{Y} &=& 2\sqrt{2}m_3^{(0)} \cos\delta\, \Delta s_{13}^{(1)} 
\label{fo11}
\\
-\frac{a\lambda_2}{X}-\frac{d\lambda_6}{Y} &=& \sqrt{2}(m_1^{(1)}-m_2^{(1)})s_{12}c_{12} \, .
\label{fo12}
\end{eqnarray}
Since we know from the data that the solar mass splitting is much smaller
than the atmospheric one, the quantities in Eq.\ (\ref{fo12})
should be smaller than naively assumed $\mathcal{O}(\lambda)$,
which yields
$m_1^{(1)}=m_2^{(1)}$ for large $s_{12}$.
{}From the point of view of the flavor model, the two contributions
to the left-hand sides of Eqs.\ (\ref{fo11})-(\ref{fo12})
should either interfere destructively or be small. The first option
represents another fine-tuning, while the second can be achieved
with a small hierarchy among the flavon vevs; $\lambda_{2,6}/\lambda_{3,4} \propto \langle \Phi_{+1}\rangle/ \langle \Phi_{\pm2}\rangle \sim\lambda$.
Irrespective of the actual origin of this feature,
it seems more appropriate to defer the discussion
of the terms proportional to $-\frac{a\lambda_2}{X}-\frac{d\lambda_6}{Y}$
to the analysis of the $\mathcal{O}(\lambda^2)$ corrections.
A similar comparison for the remaining entries of ${m}_\nu^{(1)}$ gives:
\begin{eqnarray}
\label{fo3}
-\frac{d\lambda_3}{Y}-\frac{|b|\lambda_4}{Y} &=& -2m_3^{(0)} \Delta\theta_{23}^{(1)} \, , \\
-\frac{d\lambda_3}{Y}+\frac{|b|\lambda_4}{Y} &=& m_2^{(1)}c_{12}^2+m_1^{(1)}s_{12}^2 \, ,\\
-\frac{d\lambda_3}{Y}+\frac{|b|\lambda_4}{Y} &=& m_3^{(1)} \, , \\
-\delta_a^{(1)} +\imath\alpha\frac{bd}{Y}&=& m_1^{(1)}c_{12}^2+m_2^{(1)}s_{12}^2 \, ,
\label{fo6}
\end{eqnarray}
where we chose such combinations of the 11, 22, 23 and 33 entries
that the results are particularly simple.
It may appear that the phases of $\lambda_3$
and $\lambda_4$ must be aligned so that $d\lambda_3+|b|\lambda_4$
is real up to $\mathcal{O}(\lambda^2)$ corrections. However, we still have one
phase redefinition, which we can use to impose this
condition, so it does not represent another fine-tuning.
By taking a linear combination of Eqs.\ (\ref{fo3})-(\ref{fo6}), we obtain
a consistency condition for a small solar mass splitting:
\beq
-\frac{-d\lambda_3+|b|\lambda_4-\imath\alpha bd}{Y}+\delta_a^{(1)} = (c_{12}^2-s_{12}^2)(m_2^{(1)}-m_1^{(1)}) \approx 0 \,
\label{corafo1}
\eeq
which 
determine the unphysical phase $\alpha$ and
imposes a constraint on
$\delta_a^{(1)}$, thereby
increasing the
already present fine-tuning (\ref{ft1}).
A relation of this type seems unavoidable in any neutrino mass model predicting
a degenerate spectrum.
The atmospheric mass splitting is then:
\beq
\label{amd}
\Delta m^2_\mathrm{atm} = |m_3|^2-|m_2|^2 = 4 m_3^{(0)}\mathrm{Re}[m_3^{(1)}] \, ,
\eeq
which is naturally of the order $\mathcal{O}(\lambda)$ with respect to
the neutrino mass scale and, for fixed $|\lambda_3|,|\lambda_4|$, 
it is maximal if $\lambda_3$ and $\lambda_4$
are approximately real.

Using this approach, one can also write the relations between the flavor
model and the phenomenological parameterization
at the $\mathcal{O}(\lambda^2)$ order.
These lengthy expressions, which we omit here,
give 6 independent relations between the flavor model parameters
and the variables
$m_i^{(2)}$, $\theta_{12}$, $\Delta\theta_{23}^{(2)}$ and
$s_\delta\Delta s_{13}^{(1)}$ or $c_\delta\Delta s_{13}^{(2)}$.
Hence, no further fine-tunings appear at this stage
and the solar splitting is then given by:
\begin{eqnarray}
\Delta m^2_\mathrm{sol} &=& |m_2|^2-|m_1|^2 = \nonumber\\
&=& -2m_3^{(0)}\left(\mathrm{Re}[m_2^{(2)}]-\mathrm{Re}[m_1^{(2)}]\right) \, .
\end{eqnarray}
Finally, we note that the model considered here corresponds in some
limiting cases to models already present in the literature.
Therefore, the following considerations regarding the viability of
our model and, in particular, the amount of fine-tuning necessary to
describe the neutrino oscillation data can also be applied to those
models.
For $\lambda_3=\lambda_6=0$ and real $Y_\nu$, we obtain the model
studied previously in Ref.~\cite{lmlt2}. We also note that
for $\lambda_3=\lambda_6=0$ and $X=Y$ the neutrino mass matrix (\ref{mnuour})
is identical to that considered in an $A_4$-inspired model of 
Ref.~\cite{hirsch}.

\subsection{Fine-tuning}
\label{seft}

In Section \ref{flamo}, we have seen that our model requires
a fine-tuning, necessary for arranging a small solar neutrino
mass-squared splitting. Here, we shall discuss this issue in more detail and
compare our model to other models of neutrino masses and mixing.

Addressing the issue of fine-tuning in a quantitative way
is a cumbersome task, since it inevitably requires introducing
a probability measure in the parameter space.
We shall therefore make a comparative
study, checking the performance of our neutrino mass model 
(with $\lambda_1=\lambda_5=0$)
{\it versus} another neutrino mass model which also arises
from breaking of an $U(1)$ flavor gauge symmetry
through Froggatt-Nielsen mechanism and is
regarded as rather natural. As the reference model, we chose the
HII model of Altarelli, Feruglio and Masina (AFM) \cite{afm},
which predicts a hierarchical spectrum of neutrino masses.
In order to make a comparison with another model explaining quasi-degenerate
neutrino masses, we shall also analyze the model of He, Keum and Volkas
(HKV) \cite{he} based on $A_4$ symmetry.
For completeness, 
we shall also compare our model with an {\em anarchical} seesaw
model, {\em i.e.}~one exhibiting no structure in $\YY{}$ or $M$
\cite{Hall}.

\begin{figure}
\begin{center}
\hspace{-3cm}
\includegraphics*[height=10cm]{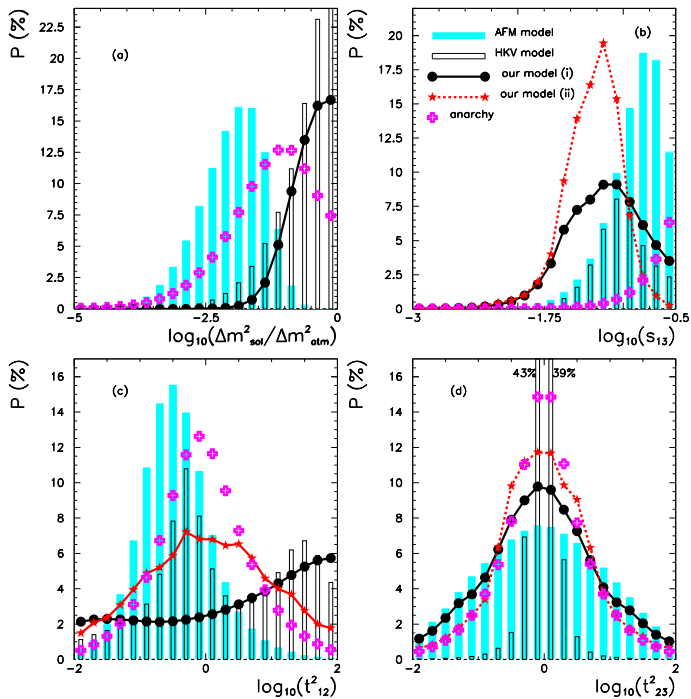}
\end{center}
\caption{Probability distributions for
$\Delta m^2_\mathrm{sol}/\Delta m^2_\mathrm{atm}$, $s_{13}$, $t^2_{12}$
and $t^2_{23}$. The lines correspond to predictions of our model for
(i) general choice of the parameters and (ii) choice of the parameters
with $\Delta m^2_\mathrm{sol}/\Delta m^2_\mathrm{atm}$
in the experimentally allowed range.
The filled (empty) histograms correspond to
the AFM (HKV) model. \label{figdis}}
\end{figure}

The comparison has been performed along the lines of the analysis
presented by AFM.
Each entry $\mathcal{O}(\lambda^n)$ allowed by symmetry
was parameterized as $fe^{\imath\omega}\lambda^n$, where $0.8\leq f\leq 1.2$
and $0\leq\omega\leq 2\pi$ were chosen randomly with a constant
probability density. We used
an optimized value $\lambda=0.35$ for the AFM model,
while we set a suggestive value $\lambda=0.22$
in our model. For the HKV model, we assumed that
all unperturbed entries are $\mathcal{O}(\lambda^0)$
and that the perturbations are $\mathcal{O}(\lambda)$ with $\lambda=0.1$.
For the anarchical model all the entries in $\YY{}$ and $M$
were assumed to be $fe^{\imath\omega}$.
We then diagonalized numerically the resulting neutrino mass
matrices and calculated four dimensionless
observables unambiguously constrained
by the present neutrino data:
$\Delta m^2_\mathrm{sol}/\Delta m^2_\mathrm{atm}$, $s_{13}$, $t^2_{12}$
and $t^2_{23}$. This procedure was repeated $10^6$ times for each model.
The resulting probability distributions of the observables
are shown in Figure \ref{figdis}.
The overall success rate of each model can be defined
as the fraction of points lying
in a four-dimensional box whose sides correspond to $3\sigma$
ranges of the observables allowed by the present data. Such a success rate
was approximately $3\times10^{-3}$ for the AFM model, $2\times10^{-2}$ for the HKV model and
$4\times10^{-4}$ for our model, the actual number depending on the
RG corrections (admitting $\lambda_{1,5}\sim\lambda_{2,6}$ does not change
this result qualitatively). 
A purely anarchical model has the success rate twice
smaller than our model.

If we consider the success rate an unambivalent measure of naturalness,
the AFM model and HKV model are favored over ours by the oscillation data.
As regards $\Delta m^2_\mathrm{sol}/\Delta m^2_\mathrm{atm}$,
the AFM distribution, peaked around $10^{-2}$ is rather wide and it could
easily account for
a wide range of values of this observable, whose experimentally allowed
$3\sigma$ range (with RG correction neglected)
is assumed with $15\%$ probability. In contrast,
the lower value of this observable in our model and in the HKV model,
the larger fine-tuning is required,
and the probability of obtaining 
$\Delta m^2_\mathrm{sol}/\Delta m^2_\mathrm{atm}$
in the allowed range is $\sim1\%$ and $\sim4\%$, respectively.
The sign of this observable in our model is positive in more than
$95\%$ of cases in our model, which justifies {\em a posteriori}
the assumptions made in Section \ref{flamo}. 
Values of $s_{13}$ come out small in all models but the anarchical one, 
with $\sim30\%$ (HKV), $\sim50\%$ (AFM) and $\sim60\%$ (our) of the
distribution in the allowed range. The atmospheric mixing is peaked
around the maximal mixing  models, with the HKV distribution being the
most narrow.

As we already argued in Section \ref{flamo}, there is no point
in discussing the solar mixing independently of the solar mass
splitting in our model, as
the consistency with experimental data introduces some
correlation between observables. As shown
in Figure \ref{figdis} (where we also plot
the distribution of conditional probability
density given that the solar-to-atmospheric ratio lies
within the experimentally allowed range),
once the fine-tuning required for the
solar-to-atmospheric mass ratio is achieved, the distribution
of $t_{12}^2$ becomes peaked around values consistent with experiment.
Similarly,
the distribution of conditional probability
density for $s_{13}$ given that the solar-to-atmospheric ratio
(empty cyan boxes) is shifted towards smaller
values of this observable, pursuant to Eq.\ (\ref{fo12}).

In conclusion, in comparison
to the AFM model and the HKV model, 
our model's overall performance is worse by a factor
of 10 to 100, following mainly from the fine-tuning necessary for the small
solar mass splitting.
However,
if our model can explain the baryon asymmetry of the Universe as
resulting from leptogenesis with a low reheating temperature, while
the AFM and HKV model cannot, this may be a hint that a
quantitatively moderate fine-tuning discussed
above
allows a glimpse at the structure of a more fundamental physics
rather than being an unnatural coincidence.

\section{Leptogenesis}

In the MSSM, the effects of supersymmetry breaking in leptogenesis
can be safely neglected. The CP asymmetries are twice larger than those
in the Standard Model and the number of channels through which the lepton
asymmetry is generated is also doubled. This is compensated by doubled
amplitudes of the washout processes and an almost doubled 
number of relativistic degrees of freedom after leptogenesis. The conversion
factors, relating the generated lepton asymmetry with the final baryon 
asymmetry are also very similar \cite{jmr}.
Therefore, the order of magnitude of 
the baryon asymmetry of the Universe resulting from leptogenesis
can be approximated by the nonsupersymmetric formula \cite{Underwood06}:
\beq
\eta_B \sim 10^{-2} \sum_{i=1}^{3} e^{1-M_i/M_1}\sum_\alpha \varepsilon_{i\alpha} \frac{K_{i\alpha}}{K_iK_\alpha} \,
\label{bmlmaster}
\eeq
where
$\alpha$ runs over distinguishable lepton flavors $\alpha=e,\mu,\tau$
(we assume a reheating
temperature $\simlt10^9\,\mathrm{GeV}$) and
$\varepsilon_{i\alpha}$ are CP asymmetries in the decays of the
right-handed neutrinos of mass $M_i$ into flavor $\alpha$.
The washout parameters are defined as
\begin{eqnarray}
K_{i\alpha}&=& \frac{\left(\Gamma(N_i\to L_\alpha H_2)+
\Gamma(N_i\to \bar{L}_\alpha H_2^\ast)\right)}{H(T=M_i)} \approx \nonumber\\
&\approx& \frac{\tilde{m}_{i\alpha}}{10^{-3}\,\mathrm{eV}}  \, ,
\label{kpar}
\end{eqnarray}
where $\tilde{m}_{i\alpha} =|{y}_{i\alpha}|^2\langle H_2\rangle^2/M_i$,
$K_i=\sum_\alpha K_{i\alpha}$, $K_\alpha=\sum_i e^{1-M_i/M_1}K_{i\alpha}$,
and
${y}$ is the neutrino
Yukawa matrix written in the basis of the mass eigenstates for the right-handed
neutrinos.
It follows from Eq.\ (\ref{kpar}) that for the light neutrinos with masses
$0.1\,\mathrm{eV}$, we need $|\varepsilon_{i\alpha}|\sim 10^{-4}$
for successful leptogenesis. However, for $M_i\simlt 10^9\,\mathrm{GeV}$
which allows avoiding the gravitino problem, a natural scale for
the CP asymmetries is
\beq
\label{natscaeps}
|\varepsilon_{i\alpha}| \sim
M_i m_\nu/\langle H_2\rangle^2\sim 10^{-6} 
\eeq
(we neglect various $\mathcal{O}(1)$ factors), 
unless, e.g., the
right-handed neutrinos are almost degenerate in masses,
since in that case the CP asymmetries in the decays of these
right-handed neutrinos are resonantly enhanced.

In our model, the right-handed neutrinos mass matrix ${M}$ in
Eq.\ (\ref{assumpt}) has two exactly degenerate eigenvalues, $M_2=M_3=Y$, which
may become slightly split by RG corrections if the scale $Q_f$ of
$U(1)$
breaking is larger than the leptogenesis scale. If the splitting
$\delta_N=1-M_2/M_3$ is much smaller than
$({y}{y}^\dagger)_{22,33}/8\pi$,
the relevant CP asymmetries are given by \cite{Pilaftsis97} (see also \cite{Covi96}):
 \beq
\varepsilon_{J\alpha} \approx
\frac{16\pi\,\delta_N\,\mathrm{Re}[{y}{y}^\dagger]_{23}\,
\mathrm{Im}[{y}_{2\alpha}{y}^\ast_{3\alpha}]}
{({y}{y}^\dagger)_{22}
({y}{y}^\dagger)_{33}({y}{y}^\dagger)_{JJ}}
\, . \label{cpass1}
 \eeq
It may appear that for $M_2=M_3$ a transformation 
$N_2\to\cos\zeta N_2+\sin\zeta N_3$ and
$N_2\to-\sin\zeta N_2+\cos\zeta N_3$ is a symmetry of the mass matrix
of the right-handed neutrinos, but it allows for rearranging the neutrino
Yukawa couplings. However, it has been noted in \cite{Turzynski04}
that if we require that the neutrino Yukawa couplings
are continuous functions of the renormalization scale
then $\zeta$ is fixed at a value corresponding to
$\mathrm{Re}[{{y}}{{y}}^\dagger]_{23}=0$.
Hence, the CP asymmetries vanish at
the scale of the exact degeneracy of $N_2$ and
$N_3$. They assume nonzero values if
$\mathrm{Re}[{{y}}{{y}}^\dagger]_{23}\neq0$
is generated through RG corrections.
In the leading order, the solutions of
the RG equations for degenerate right-handed
neutrinos are:
\begin{eqnarray}
\delta_N &\approx& 4
\left(
({y}{y}^\dagger)_{22}-
({y}{y}^\dagger)_{33}\right)
\Delta t = \nonumber\\
&=& 8\langle H_2 \rangle^{-2} |b\lambda_4^\ast+d\lambda_3|
\Delta t \, , \label{solrgedelta} \\
\mathrm{Re}[{y}{y}^\dagger]_{23}
&\approx&
\mathrm{Re}[{y}_{23}{y}^\ast_{33}] y_\tau^2\Delta t= \nonumber\\
&=& \langle H_2 \rangle^{-2}
\,\frac{|b|d\,\mathrm{Im}[\lambda_3\lambda_4]}{|b\lambda_4^\ast+d\lambda_3|}\, y_\tau^2\Delta t \, ,
\label{solrgere}
\end{eqnarray}
where $y_\tau\sim 10^{-2}\tan\beta$ is the
tau Yukawa coupling and
$\Delta t=(4\pi)^{-2}\ln(Q_f/Y)\sim0.1+0.006\ln(Q_f/10^7Y)$.
Other combinations of parameters appearing in (\ref{cpass1})
receive negligible RG corrections and can be replaced by their
values at the scale $Q_f$. The CP asymmetries can be then expressed as:
 \beq
\varepsilon_{2\alpha}=\varepsilon_{3\alpha}
\approx \frac{64\pi \,y_\tau^2\,|b|d
\,\mathrm{Im}[\lambda_3\lambda_4]\,(\Delta
t)^2}{\left(\frac{|b|^2+d^2}{2}\right)^3} \times \left\{
\begin{array}{l}
\mathcal{O}(\lambda^4) 
\\
-|b|^2 
\\
d^2 
\end{array}
\right.
\label{ourcp}
 \eeq
where the upper, middle and lower factors correspond to $\alpha=e,\mu,\tau$,
respectively.
The other CP asymmetries,
$\varepsilon_{1\alpha}$, are much smaller and can be neglected.
We estimate from (\ref{ourcp})
that $|\varepsilon_{2\alpha}|\simlt 10^{-6}\tan^2\beta$,
so the CP asymmetries are sufficiently large for successful leptogenesis
if $\tan\beta\simgt10$.

\begin{figure}
\begin{center}
\hspace*{1cm}
\includegraphics*[height=12cm]{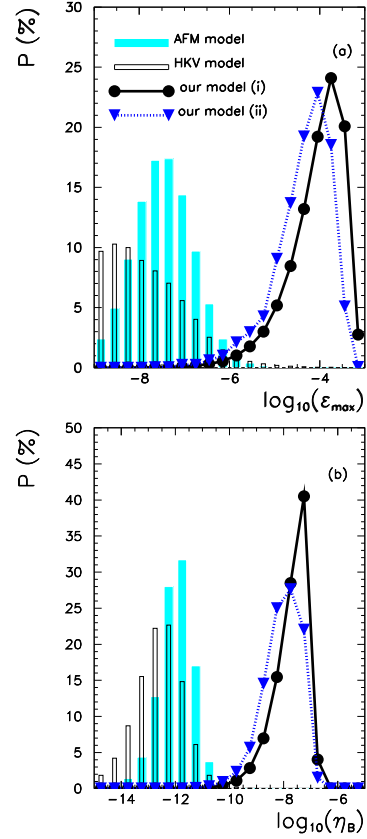}
\end{center}
\caption{Probability distributions for
$\varepsilon_\mathrm{max}$ and $\eta_B$.
The lines correspond to predictions of our model for
(i) general choice of the parameters and (ii) choice of the parameters
satisfying constraints from the oscillations data.
The filled (empty) histograms correspond to
the AFM (HKV) model. \label{figeps}}
\end{figure}

The predictions of our neutrino mass model for the CP asymmetries
given in Eq.\ (\ref{ourcp}) are presented in Figure \ref{figeps}, where
we plot the distribution probability for
$\varepsilon_\mathrm{max}=\mathrm{max}\{|\varepsilon_j|;\,j=1,2,3\}$, where
$\varepsilon_j=\sum_\alpha\varepsilon_{j\alpha}$.
The numerical procedure is identical to that described in Section \ref{seft},
with the exception that we scan over $10^7$ points in the parameter space
for the conditional probability distribution.
For definiteness, we assume that $\Delta t=0.1$, $\tan\beta=10$
and $M_1=10^9\,\mathrm{GeV}$.
The black, solid lines correspond to the general parameter
choice in our model, while the blue, dotted lines represent
parameter choices satisfying all four
phenomenological constraints; the filled (empty) histogram shows
the predictions of the AFM (HKV) model.
In this Figure, we also show the estimates the baryon
asymmetry of the Universe $\eta_B$ given by formula (\ref{bmlmaster}).
According to the discussion in Section \ref{flamo}, a requirement of a
small solar neutrino mass splitting favors almost real 
$\lambda_3$ and $\lambda_4$,
which in turn leads to a certain suppression of
the CP asymmetry (\ref{ourcp}).
Moreover, the fact that $\varepsilon_{j\mu}$ and $\varepsilon_{j\tau}$
are proportional to $-|b^2|$ and $d^2$, respectively, is the reason for
another slight suppression of the resulting baryon asymmetry.
These suppressions can, however, be easily overcome by the enhancement
of the tau Yukawa coupling for $\tan\beta\simgt10$, and
we conclude that we can easily have
the CP asymmetry of the order of $10^{-4}$ which can account for
the baryon asymmetry of the Universe.
As regards the AFM and HKV models, with an optimistic assumption
that the largest values of $\varepsilon_\mathrm{max}$ correspond
to washout as small as $0.1$, we conclude that these models can
be only marginally consistent with baryogenesis via leptogenesis
for a low reheating temperature.

The crucial ingredient of our model which allows for a low-scale
leptogenesis is the assumption that the $L_\mu-L_\tau$ flavor symmetry
is broken at a scale $M_X$ much larger than the leptogenesis scale $M_{N_A}$
and that this breaking is transmitted to the mass matrix of the 
right-handed neutrinos only through RG corrections. At the leptogenesis scale,
the masses of the pseudo-Dirac right-handed neutrinos are then split by
a factor proportional to the neutrino Yukawa couplings and it is precisely
this small splitting which makes it possible to overcome the naive scaling
(\ref{natscaeps}) \cite{Turzynski04}.
For comparison, in model described in 
\cite{lmlt2}, also based on $L_\mu-L_\tau$ symmetry, the scale
of symmetry breaking is identified  with the leptogenesis scale, 
the RG corrections are absent
and the pseudo-Dirac right-handed neutrinos are exactly degenerate, which leads
to a vanishing CP asymmetry in their decays. (Besides, only one CP violating phase is assumed in this model; although it is straightforward to include other phases in the neutrino sector, this would not change the latter conclusion.) Hence, the scaling (\ref{natscaeps}) holds approximately and large values of the right-handed neutrino masses of are necessary for successful leptogenesis.
In contrast, in the model of \cite{lmlt1a} (in which the symmetry breaking scale is also identified with the leptogenesis scale), the $L_\mu-L_\tau$ symmetry
is broken in {\em both} the neutrino Yukawa matrix and the mass matrix of
the right-handed neutrinos which again leads to the approximate scaling
(\ref{natscaeps}) and large values of the right-handed neutrino masses
necessary for successful leptogenesis.

\section{Conclusion}

In this work, we have considered a
neutrino mass model where the neutrino Yukawa and mass structures are dictated by
the flavor symmetry, $L_\mu-L_\tau$, and its breaking patterns which is controlled by
an additional discrete symmetry.
The model requires a fine-tuning to correctly predict the smallness of
the solar mass splitting. Taking the expansion parameter $\lambda=0.22$,
we have made a quantitative discussion on
the fine-tuning in the combined explanation of  all the low-energy observables and
the demanded baryon asymmetry of the Universe. Once such a fine-tuning is ensured, the bi-large pattern
of mixing angles and a successful leptogenesis
with a low reheating temperature becomes a natural prediction
of this model.  In addition, the Dirac CP phase is generically order-one while
the reactor mixing angle $\theta_{13}$ is peaked at $\theta_{13}=0.06$.

\begin{acknowledgments}
KT thanks S.~Pokorski and P.~H.~Chankowski for
discussions at early stages of this project. 
This work is supported by the Department of Energy.
\end{acknowledgments}

\end{document}